\documentclass[3p,12pt, times]{elsarticle}
\usepackage{amsmath}
\usepackage{amssymb}
\usepackage{algorithm}
\usepackage[noend]{algpseudocode}
\usepackage{graphicx}
\usepackage{multicol}
\usepackage{multirow}
\usepackage{float}
\usepackage{array}
\usepackage{siunitx}
\usepackage{caption}
\usepackage{subcaption}
\usepackage{adjustbox}
\usepackage{pifont}
\newcommand{\cmark}{\ding{51}}%
\newcommand{\xmark}{\ding{55}}%
\begin{document}

\begin{frontmatter}

\title{Towards Agent-Based Model Specification of Smart Grid: A Cognitive Agent-Based Computing Approach}
\author[labela]{Waseem Akram} 
\author[labela]{Muaz A. Niazi$^*$}
\author[label]{Laszlo Barna Iantovics}
\address[labela]{Computer Science Department,\\ COMSATS Institute of Information Technology,\\Islamabad, Pakistan}
\address[label]{Corresponding author}

%

\begin{abstract}
A smart grid can be considered as a complex network where each node represents a generation unit or a consumer. Whereas links can be used to represent transmission lines. One way to study complex systems is by using the agent-based modeling (ABM) paradigm. An ABM is a way of representing a complex system of autonomous agents interacting with each other. Previously, a number of studies have been presented in the smart grid domain making use of the ABM paradigm. However, to the best of our knowledge, none of these studies have focused on the specification aspect of ABM. An ABM specification is important not only for understanding but also for replication of the model. In this study, we focus on development as well as specification of ABM for smart grid. We propose an ABM by using a combination of agent-based and complex network-based approaches. For ABM specification, we use ODD and DREAM specification approaches. We analyze these two specification approaches qualitatively as well as quantitatively. Extensive experiments demonstrate that DREAM is a most useful approach as compared with ODD for modeling as well as for replication of models for smart grid.
 
\end{abstract}

\begin{keyword}
Agent-based modeling \sep Complex networks \sep Smart Grid
\end{keyword}
\end{frontmatter}
\section{Introduction}
\label{sec:intro}
A smart grid focuses on the complex interactions between utility service providers and consumers. It involves the non-linear dialogue of power and information data between utility service providers and consumers \cite{ellabban2016smart}. The complex interactions in the form of repeated auction, fluctuating supply and demand add complexity to the nature of a smart grid. Because of this complex nature, it can be considered as a complex system.

The study and understanding of any complex system are associated with the modeling of the system. Modeling complex system allows better understanding and analyzing the emergent behavior of each entity involved in the system \cite{epstein2008model}. Being a complex system, a smart grid can also essentially be modeled in the form of either agent-based or complex network-based models \cite{niazi2017towards,niazi2013complex}. These models can well represent the smart grid in term of its various components, their behavior, and communication among them for the energy distribution and management.
 
A particular way of modeling smart grid as a complex network is by including its various components such as generating units, consumers, distributors and other components as nodes and communication as edges. Chasing et al. have developed the complex network model for the US power grid by considering the number of nodes as power sources and consumers, while the number of edges as communication lines \cite{chassin2005evaluating}. After developing different complex networks, we able to use the mathematical tool for computing centrality measures and metrics on such networks. Then these measurements will allow studying the global behavior of each component in large-scale power system network.

Regarding complex systems, Batool and Niazi \cite{batool2017modeling} have proposed a hybrid modeling approach for the Internet of things (IoT) domain. In this work, a combination of agent-based and complex network-based approaches are used for modeling IoT-based complex scenarios. This work demonstrates the use of standard complex networks such as small-world, scale-free and random network. However, this previous work is not extended and applied to the smart grid domain. We think this previous modeling methodology is also useful for the smart grid-based complex scenarios. 

In scientific literature, agent-based modeling (ABM) and the multi-agent system (MAS) are effectively used in a smart grid domain. Some of these works have been discussed in the later section of the paper (See Discussion section). However, these works lack in any ABM specification approach for documenting ABM. An ABM specification is most important for understanding as well as replicating ABM. So there is a need for an easily understandable methodology to describe an ABM, specifically in the smart grid domain. 

The purpose of this study is twofold. First, we propose an ABM for the smart grid using a combination of agent-based and complex networks-based approaches. In our work, we will use the previously developed standard complex networks such as small-world, scale-free and random network. For validation, we also present a hybrid centrality-based routing algorithm. This will allow the end to end delivery between consumers and utility service providers. Second, for ABM specification, we follow two specification approaches, the one is ODD (short for Overview, Design concept, and Details) \cite{grimm2006standard} and the second is a DREAM (short for Descriptive Agent-based modeling) \cite{niazi2017towards}. Then we present a comparative analysis of both specification techniques.

Our main contributions can be listed as follows:
\begin{enumerate}
    \item A proposed approach for modeling and simulation of smart grid using complex network and agent-based modeling approaches.
    \item A hybrid centrality-based routing algorithm.
    \item The ODD specification approach used for ABM model of smart grid.
    \item The DREAM specification approach used for ABM model of smart grid.
    \item A comparative analysis of ODD and DREAM specification techniques.
\end{enumerate}

The rest of the paper is structured as follows: Section 2 presents basic background and concepts, in Section 3 a model development is presented, Section 4 is dedicated for results and discussions, the paper ends with conclusions formulated in Section 5.

\section{Background}
\label{sec-background}
In this section, we present basic concept and understanding of cognitive agent-based computing approach, DREAM, and ODD specification approaches. 
\subsection{Cognitive agent-based computing approach}\label{sec:cabc}
Niazi and Hussain in \cite{niazi2012cognitive} presented a unified cognitive agent-based computing approach for agent-based and complex network-based modeling. This approach involves the process of taking any complex system from the real world and to convert it into a suitable simple model by using specific modeling level such as exploratory or descriptive agent-based approach. The exploratory approach involves the use of agents to explore the complex system, identifies which agent-based model is feasible for the specific problem, then develops the proof concept and also explains what kind of data is required for validation and verification of the model. The Descriptive agent-based model level is the process of presenting ABM specification in the form of pseudo-code, a complex network of the model and social network analysis. Other modeling levels involve complex network and validation/ verification of the model.
\subsection{Descriptive agent-based modeling}\label{sec:dream}
\begin{figure}[h]
\begin{center}
\includegraphics[width=5.0 cm, height=6.0 cm]{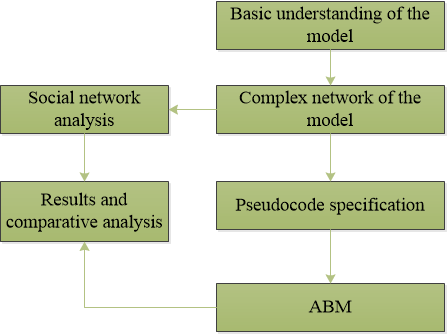}
\caption{DREAM methodology for ABM specification adopted from \cite{niazi2017towards}. We can see that it starts from a basic understanding of the model, then followed by developing a complex network of the model. There are two steps after network formation, one is to present pseudocode specification, this step allows for the actual code translation, second after network formation is by applying social network analysis tool to compute centrality measures of the network. The final step involves the results analysis.}\label{fig:dream}
\end{center}
\end{figure}

Descriptive agent-based modeling (DREAM) is a cognitive agent-based computing approach developed by Niazi in \cite{niazi2017towards}. DREAM offers an ABM specification technique which comprises of developing a complex network of the ABM, pseudo-code specification model and social network analysis of the network model. It offers a detailed description of ABM as well as visual based analysis. It provides an easy translation of network model into pseudo-code followed by ABM development. In figure \ref{fig:dream}, Dream methodology has been shown.
\subsection{Overview, Design concept and Details}\label{sec:odd}
\begin{figure}[H]
\begin{center}
  \includegraphics[width=8.0 cm, height=10.0 cm]{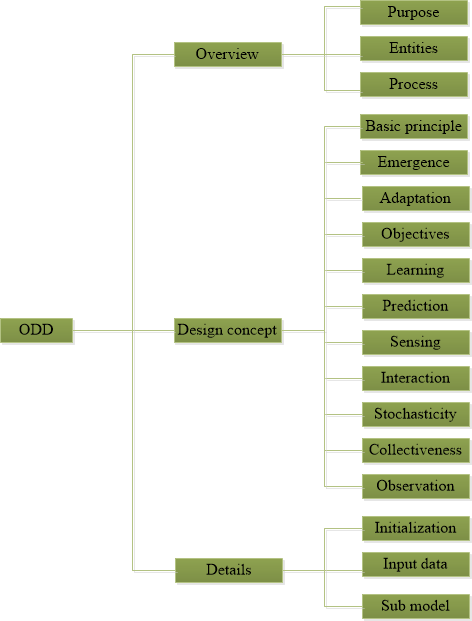}
  \caption{ODD methodology for ABM specification adopted from \cite{grimm2010odd}. We can see that ODD is divided into three main sections that are named: overview, design concept, and details. Then each section is further divided into subsections. These sections cover key features of the model.}\label{fig:odd}
\end{center}
\end{figure}

Overview, Design concept and Details (ODD) is originally developed by Grimm et al. in \cite{grimm2006standard} and then an updated version is presented in \cite{grimm2010odd}. It is a textually based specification technique for documenting ABM. It provides a checklist which covers key features of the model. It comprises of three main sections which are Overview, Design concept, and Details. These sections are further divided into subsections. In figure \ref{fig:odd}, ODD specification methodology is presented. A detailed description can be found in \cite{grimm2010odd}.
\section{Model development}
\label{sec-model development}
In this section, we present a case scenario of the smart grid, followed by ODD and DREAM specification of the model.

\begin{figure}[H]
\begin{center}
\includegraphics[width=8.0 cm, height=6.0 cm]{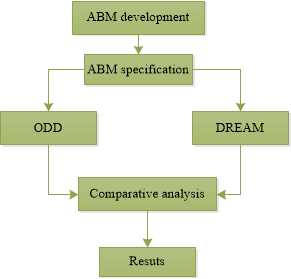}
\caption{Our research method. We can see that the first step comprises of ABM development followed by the specification of the ABM. Two specification methods are adopted (ODD and DREAM). After specification, the next step is to compare and analyze both specifications approaches qualitatively as well as quantitatively. The final step shows results of the comparative analysis.}\label{fig:researchmethod}
\end{center}
\end{figure}

Figure \ref{fig:researchmethod} shows our research methodology. First, we develop ABM for the smart grid, then followed by two specification approaches ODD and DREAM. After this, we perform a comparative analysis of these specification approaches. In the final step, we present results.
\subsection{Scenario of smart grid}\label{sec:scenrioofSG}
To model smart grid, let us consider a large set of network, in which different consumers and power generation units are connected with each other through different configuration. To model different possibility of different configurations of the large-scale power system, we use standard complex networks such as small-world \cite{watts1998collective}, scale-free \cite{barabasi1999mean} and random network \cite{barrenechea2004lattice}. For validation, we apply routing techniques such as random walk and centrality-based routing.

The routing process involves the selection of path from the source toward the destination. Routing strategy in complex networks can be categorized into two types, i.e. Local and Global Routing. The local routing strategy needs local information about neighbor nodes. These are including local static routing, local dynamic routing, and local pheromone routing \cite{kleineberg2017collective}. On the other side, global routing strategy needs global information like topological structure, characteristics of each node and real-time information. These are including shortest path routing, efficient routing, and global dynamic routing \cite{lin2016advanced}.

For large-scale complex networks, global routing remains problematic. It is difficult to have the characteristic of each node and to have real time information. Another difficulty consists in the increases of computational time. While on the other hand, local routing remains promising for large scale real world complex networks. It offers less computational time as well as easy implementation.

In a smart grid environment, two types of routing occur. The one is energy demand from consumers side to generation unit, while the generation unit responses by providing energy to consumers. The second is data and information routing about demand profile from consumers side and energy cost from grid unit \cite{rekik2016geographic}. 

\subsection{Model specification according to ODD}\label{sec:msfo}
\subsubsection{Overview}
\begin{enumerate}
\item The purpose of the model:\\
To understand how a combination of agent-based and complex network-based modeling approaches can be used to simulate large-scale power system. Further, how routing techniques can be used to validate the model.

\item The involved entities:\\
The model consists of three types of agents that have the names; consumers, producers, and walkers that are represented by nodes in a complex network. The model allows that producers and consumers are generated randomly in a network. Producers generate power and can transmit to the consumers through communication lines that we called links. Consumers demand and finally use energy power. The environment is set as a complex network where nodes represent producers and consumers and links represent transmission lines. State variables visited? and consumer? are used to mark once a node being visited and to check is any available consumer? node in the neighbor list.

\item Routing purpose:\\
For routing purpose, the concept of walkers is deployed. Initially, the walkers are located at the producer's nodes. They search for the neighbor nodes and move to one of the neighbor's node. Once a node is visited, it is marked as visited?. The walkers also check for the consumer's node. The simulation time is kept as continues. At each time step, plots are generated in order to measure visited nodes and visiting consumers.
\end{enumerate}
\subsubsection{Design concept}
\begin{enumerate}
\item Basic principles:\\
The basic hypothesis of our model is that a cognitive agent-based computing approach is better for modeling and simulation of the large-scale power system. In our approach, we used a combination of agent-based and complex network based modeling approaches. We used previously developed complex network models such as small-world, scale-free and random network to simulate a smart grid based environment.
 
\item Emergence:\\
The ``emergence'' feature shows information about ``What kind of outputs of the model are modeled''. In other words, we can say that what are the expected results from the model. In case of our approach, the routing techniques (random walk and centrality routing) are used for transmission from producers towards consumers. The key results are the computation of end to end delivery from producers towards consumers.
 
\item Adaptation:\\
Adaptive feature of the model shows decision making capability for the agents against the changing environment. Decisions are taken by using well-defined constraints to adapt the variation in the environment accordingly. There are two rules applied for making a decision. When using random walk, the walker search for neighbor's node and select one of them, while by using centrality routing, the walkers search for a neighbor node with maximum value and select that one.
  
\item Objectives:\\
In an adaptive environment, individual agents also receive effects or rewards from the environment for their adaptive behavior to achieve one's own objective. In our model, the main objective is to measure how much time is taken while moving from one node to the other. 
 
\item Sensing:\\ 
In the decision-making process among agents, there are some specific features related to each agent which allow communicating neighbors to make their decision according to values of those features. In our case scenario, the walkers use the sensing property, if a neighbor node is already been visited then they avoid rerouting. They also sense for consumers, if any visited node is a consumer, then they deliver packets or energy. 

\item Interaction:\\ 
Producers and consumers can communicate with each other for power transmission. 

\item Stochasticity:\\ 
The routing process is modeled as random.

\item Observation:\\ 
When the simulation is running, at each time step the following data are collected.
\begin{enumerate}
\item Number of nodes.
\item Number of producers.
\item Number of consumer.
\item Number of walkers.
\item Number of nodes with visited.
\item Number of consumers with visited.
\end{enumerate}
\end{enumerate}
\subsubsection{Details}
The details section of the ODD protocol covers features of the model about what is the initial state of the model, what kind of data is used, and what types of parameters and parameter values are set in the model. 
\begin{enumerate}
\item Initialization:\\
The model is implemented using NetLogo an agent-based modeling tool. The model environment is initialized by calling ``draw-network'' method. This method is used to draw any selected network. Then consumers and producers are generated randomly by specifying their number. After this, the walkers are placed at the producer's location.  

\item Input data:\\ 
The standard complex networks are generated and kept as external source files. These network files are used as input for the model.

\item Submodels:\\ 
The model parameters and parameter values are given in table \ref{tbl:metrics}.

\begin{table}[H]
\begin{center}
\caption{Evaluation metrics: These parameters and parameter values are used for model simulation. The region shows the simulation environment which is kept as 100 by 100. The number of nodes in the network is considered as 500, 800, 1000. The number of power sources and consumers are used as 5, 10, 50, 100, 150, 200. Three different standard complex networks are used. For routing purpose, random walk and centrality-based routing algorithm are applied. The Performance of the model is measured in term of average delivery rate (computation time from sources to the consumers). A series of experiments were carried out during simulation model.}\label{tbl:metrics}
\begin{tabular}{l | l}
\hline
\textbf{Parameter} & \textbf{Value} \\ \hline
Region & $ 100 X 100 $ \\ \hline
No. of Nodes & 500, 800, 1000 \\ \hline
Power sources & 5, 10, 50, 100 \\ \hline
Power consumer & 50, 100, 150, 200 \\ \hline
Network & small-world, scale-free, random network \\ \hline
Routing & random-walk, centrality-based routing \\ \hline
Performance measure & Average delivery rate \\ \hline
No. of runs & $ 4 (1, 10, 20,30 runs)$ \\ \hline
\end{tabular}
\end{center}
\end{table}
\end{enumerate}

\subsection{Model specification according to DREAM}\label{sec:msfd}
In this section, we present our ABM documentation according to DREAM specification approach. We describe our model using Pseudo-code specification part of the DREAM approach.
\subsubsection{Agent design}
There are two types of agents which are used in the simulation model. \emph{Agents by node type}, in our simulation model, we used complex networks. These complex networks consist of nodes which are connected through communication lines called links. These node agents represent producers and consumers in the network. \emph{Agents by walker type}, for routing purpose the Walker concept is deployed. These walkers initially placed on producers node. They have the ability to move around the network.

\begin{algorithm}[H]
\caption{Breed \textbf{Node:} This node agent is used to represent producers and consumers in the network}
\emph{Internal Variables:} $<source?, target?, visited?>$
\begin{algorithmic}[1]
	 \State \textbf{source?:} All nodes that represent sources(used for producers)
	\State \textbf{target?: } All nodes that represent target (used for consumers)
	\State \textbf{visited?:} Used to check the status of a Node agent
\end{algorithmic}
\label{algo:breed-node}
\end{algorithm}
In the Node specification model, first, we described the Breed Node agent. The ``Breed'' is a global keyword in NetLogo(Agent-based modeling toolkit) describing a set of similar-behavior agents. As we noted first that nodes are used in the network to represent producers and consumers. After this, we specified the internal variables for Node agent. There are three internal variables used for Node agent. The source? this variable is used to represent producers or generating unit in power system. The target? variable is used to represent consumers in a power system environment. The last one visited? is used to check the status of the node agent where it is visited or not.
Next, we define a specification for Walker agent.
\begin{algorithm}[H]
\caption{Breed \textbf{Walker:} This agent is used for walker that can move around the network}
\emph{Internal Variables:} $<location, is$-$finish?, location$-$list?>$
\begin{algorithmic}[1]
  \State \textbf{location:} Keep current location information of the walker
  \State \textbf{is$-$finish?: } Check the finish goal
  \State \textbf{location$-$list:} Keep all visited locations record
\end{algorithmic}
\label{algo:breed-walker}
\end{algorithm}
The breed Walker represents Walker agents. These walkers are deployed for routing purpose that can move around the network. Next, we specified three internal variables for Walker agent. The first one is location. The location variable is used to keep information about the current location of a walker. The is-finish? is a Boolean variable that returns true if the finish condition is met. The location-list is a list variable that used as a memory with a walker. This variable keeps information of all visited locations. 
After describing agent specification model, next, we are going to present global variables specification model.

\subsubsection{Global}
For setup simulation, the key variables are five input global variables. 
\begin{algorithm}[H]
\caption{\textbf{Input Globals:} $<get$-$network$-$type, num$-$node, num$-$walker, num$-$source, num$-$target>$ }
\begin{algorithmic}[1]
\Statex \emph{Chooser:}
	\Statex \hspace{\algorithmicindent} \textbf{get$-$network$-$type:} Used for selecting network type from the giving list
\Statex \emph{Slider:}
 \Statex \hspace{\algorithmicindent} \textbf{num$-$node:} Used for specifying number of nodes in a network
	\Statex \hspace{\algorithmicindent} \textbf{num$-$walker:} Used for specifying number of walker in a network
	\Statex \hspace{\algorithmicindent} \textbf{num$-$source:} Used for creating number of producer nodes
	\Statex \hspace{\algorithmicindent} \textbf{num$-$target:} Used for creating number of consumer nodes
\label{algo:Input-globals}
\end{algorithmic}
\end{algorithm}
Here, five input global variables are used. The get-network-type is input provided by Chooser (GUI element of the NetLogo simulation toolkit). This is used for selecting network type from the available list. The network list comprises of the small-world, scale-free and random network. The other four input variables are provided by Slider (GUI element of NetLogo). The num-node is used to specify the number of nodes in the network. The num-walker is used to specify the number of Walker in the network. The num-source and num-target are used for specification of a number of producer and consumer respectfully.

\subsubsection{Setup procedure}
Here, we present the main setup procedure for model initialization.

\begin{algorithm}[H]
\caption{Procedure \textbf{setup:} Creating simulation environment}
\emph{Input:} All global parameters from the user interface\\
\emph{Output:} Setup simulation environment\\	 
\textbf{begin}
\begin{algorithmic}[1]
	\State clear screen
    \State draw-selected-network
    \State add-sources
    \State add-targets
    \State calculate-centrality
\end{algorithmic}
\textbf{End}
\label{algo:pro-setup}
\end{algorithm}
The setup procedure is the global simulation setup specification model. This is used to create a simulation model. The input parameters are provided by a user interface. The procedure starts with the calling of clear-all function. It will clear all the previous work. Next, all four individual procedures are called. The first one is ``draw-selected-network''. Next, we describe these individual procedures.
\begin{algorithm}[H]
\caption{Procedure \textbf{draw-selected-network:} Creating the desired network}
\emph{Input:} network-file, list of available network\\
\emph{Output:} Setup the selected network\\	 
\textbf{begin}
\begin{algorithmic}[1]
	\If{network-type $=$ "small-world"}
      \State [draw-small-world]
    \Else
    \If{network-type $=$ "scale-free"}
      \State [draw-scale-free]
   \Else {network-type $=$ "random-network"} 
      \State [draw-random-network]
     \EndIf
   \EndIf
\end{algorithmic}
\textbf{End}
\label{algo:pro-draw-selected-network}
\end{algorithm}
``draw-selected-network'' is a procedure which is used to setup the selected network from the giving options (a list of networks). The procedure checks the input type and called the appropriate function of it. Here three individual procedures are used. Next, we present the specification model of these individual procedures which are called by draw-selected-network.

\subsubsection{Small-world network}
In this network topology, any individual is linked to any other individual by a maximum of six edges. In social network terminology, this is called ``six degrees of separation''. It has fewer nodes with more links. It is satisfied by: $ G(V,E), L\propto \log(N) $.
\begin{algorithm}[H]
\caption{Procedure \textbf{draw-small-world:} Creating small-world network}
\emph{Input:} N\\
\emph{Output:} Setup small-world network
\begin{algorithmic}
\Statex create-nodes \emph{N};
\ForAll {Nodes};
  \Statex set all-wired? false;
  \While { all-wired? $!=$ true}
		\State wired-them [connect-nodes];
    	\State set all-wired=\textbf{Do-calculation [\textbf{clustering-coefficient}]};
		\State \textbf{Find}clustering-coefficient;
     	\State \textbf{If} no Node left\textbf{then} stop;
  		\EndWhile
\EndFor
\textbf{end}
\State create link with one node to other
\State calculate shortest path [set node i=1;
\State  \textbf{count} distance of node i to every other node;
\State  i++;
\State  set shortest path=\textbf{min} distance]
\State set layout circle;
\end{algorithmic}
\textbf{End}
\label{algo:pro-draw-small-world}
\end{algorithm}

\subsubsection{Scale-free network}
In this network, the link of an individual with another individual is uneven. There are some nodes which have dense connections, while some others have fewer connections. The dense connections are also called hubs. These hubs have the tendency to join with other new nodes. This network follows the power law of the degree distribution. The probability of joining new nodes with the existing hub can be defined by:
\begin{equation}
\prod_(ki)=\frac{K_i}{\sum_{j} Kj}
\label{euation:forscalefree}
\end{equation}
In Eq.\ref{euation:forscalefree}, $K_i$ represents degree of hub $i$.
\begin{algorithm}[H]
\caption{Procedure \textbf{draw-scale-free:} Creating scale-free network}
\begin{algorithmic}
\State \emph{Input:} N
\State \emph{Output:} Setup scale-free network
\State set shapes circles;
\State create node 1;
\State set node=nobody;
\State create node 1;
\State $[$ set node = new-node$]$
\If {old-node = nobody}
	\State [create link with old-node];
\While{\textbf{count} nodes $<$ total-nodes}
		\State [add nodes];
	\EndWhile		
\EndIf
\State set layout spring;
\State set layout circle;
\State \textbf{End}
\end{algorithmic}
\label{algo:pro-draw-scale-free}
\end{algorithm}

\subsubsection{Random network}
In this network, each individual node is formed randomly; there is no specific structure to be followed. This network can be formed by joining vertex with other arbitrary. Formally a random network $G_R(N,P)$ is framed with edges associated with likelihood, p given that $0 < P < 1$ . The connectivity of nodes with other does not depend on the degree of nodes.
\begin{algorithm}[H]
\caption{Procedure \textbf{draw-random-network:} Creating random network}
\begin{algorithmic}
\State \emph{Input:} N
\State \emph{Output:} Setup random network
\State create nodes \emph{N};
\ForAll {Nodes}
	\State set shapes circles;
	\State set location random location;
	\State create-links with nodes with p;
	\State check p $> 0$ and p $< 1$;
	\EndFor
\State \textbf{End}
\end{algorithmic}
\label{algo:pro-random}
\end{algorithm}
Next, we present all other procedures that describe different processes which are called by the main setup procedure.

\subsubsection{Setup source and target nodes procedures}
``add-source'' procedure is used to setup source nodes (producers) on the network. It takes network type, nodes, links, and a number of the source from the user interface. Next create source nodes of a specified number from user randomly on the network. 
\begin{algorithm}[H]
\caption{Procedure \textbf{add-source:} Setup sources on the network}
\begin{algorithmic}
\State \emph{Input:} network-type, num-source
\State \emph{Output:} Setup source nodes on the network
\State \textbf{begin}
 	\State \hspace{\algorithmicindent} random nodes[;
	\State \hspace{\algorithmicindent} set source true;
\State \textbf{End}
\end{algorithmic}
\label{algor:add-source}
\end{algorithm}
``add-target'' procedure is used to setup target nodes (consumers) on the network. It takes network type, nodes, links, and a number of the target from the user interface. Next create target nodes of a specified number from user randomly on the network.
\begin{algorithm}[H]
\caption{Procedure \textbf{add-target:} Setup target on the network}
\begin{algorithmic}
\State \emph{Input:} network-type, num-target
\State \emph{Output:} Setup target nodes on the network
\State \textbf{begin}
 	\State \hspace{\algorithmicindent} Select random nodes[;
	\State \hspace{\algorithmicindent} set target true;
\State \textbf{End}
\end{algorithmic}
\label{algor:add-target}
\end{algorithm}
After setting source and target nodes on the network. Next, the main setup procedure calls ``calculate-centrality'' procedure. Next, we present calculate-centrality procedure.

\subsubsection{Centrality measure}
The complex network provides centrality measure feature. The centrality measure is the process to find out which vertex is the most central which is important. It is widely used for measuring the relative importance of nodes within a network. It is a numerical number assigned to each node for pair-wise comparison within a whole network. It has four types.
\begin{enumerate}
\item Degree of nodes:\\
It measures the total number of connections that a particular node has in a network. A node with a higher degree has most importance as compared with those which have a lower degree. If a node with a higher degree is removed, then it can disrupt the structure as well as the flow of the network. 

\begin{algorithm}[H]
\caption{Procedure \textbf{calculate-degree}: Calculates degree centrality on the network}
\begin{algorithmic}
\State \emph{Input:} selected network
\State \emph{Output:} calculates degree centrality for all nodes
\ForAll {Nodes}
	\State set degree \textbf{count} all connected links
    \State set label degree
    \EndFor
\State \textbf{End}
\end{algorithmic}
\label{algor:degree}
\end{algorithm}

\item Closeness of nodes:\\
It is used to find out how much data from a particular node moves to every other node in a network. To calculate closeness of a particular node $i$ from all other nodes $t$. Mathematical it can be written as:
\begin{equation}
C_{closeness(s)}=\sum \frac{1}{dist(i,t)}
\end{equation}
\begin{algorithm}[H]
\caption{Procedure \textbf{calculate-closeness}: Calculates closeness centrality on the network}
\begin{algorithmic}
\State \emph{Input:} selected network
\State \emph{Output:} calculates closeness centrality for all nodes
\ForAll {Nodes}
	\State set closeness $ C_{closeness(i)}=\sum{\frac{1}{dist{(i,t)}}}$
    \State set label closeness
    \EndFor
\State \textbf{End}
\end{algorithmic}
\label{algo:closeness}
\end{algorithm}

\item Betweenness of nodes:\\
Betweenness centrality is the process of counting the number of times a specific vertex comes in the shortest path between any two vertexes in a network. It has the capability to observe the network transmission. Mathematically can be written as:
\begin{equation}
C_{betweenness(i)}=\sum{\frac{\sigma st(i)}{\sigma st}}
\label{eq:calBetweenness}
\end{equation}
Where $\sigma st(i) $ denotes the number of shortest paths between nodes $s$ and $t$ passing through the node $i$. While $ \sigma st $ is the total number of shortest paths exist between node $s$ and $t$. 
\begin{algorithm}[H]
\caption{Procedure \textbf{calculate-betweenness}: Calculates closeness centrality on the network}
\begin{algorithmic}
\State \emph{Input:} selected network
\State \emph{Output:} calculates betweenness centrality for all nodes
\ForAll {Nodes}
	\State set betweenness $ C_{betweenness(i)}=\sum{\frac{\sigma st(i)}{\sigma st}}$
    \State set label betweenness
    \EndFor
\State \textbf{End}
\end{algorithmic}
\label{algo:betweenness}
\end{algorithm}

\item Eigen-vector of nodes:\\
It measures the impact of a particular node in a network. It defines which node is connected to the most important node in a network. It depends on neighbors in term of connections that neighbors have with other nodes in a network. 
\begin{algorithm}[H]
\caption{Procedure \textbf{calculate-eigenvector}: Calculates eigenvector centrality on the network}
\begin{algorithmic}
\State \emph{Input:} selected network
\State \emph{Output:} calculates eigenvector centrality for all nodes
\ForAll {Nodes}
	\State set eigenvector $ C_{eigenvector(i)}=\frac{1}{\lambda}{\sum_{t \in g}^{}A_{adj-matrix}} * X_t \,\,\,\,\,\, OR \,\,\,\,\,\, Ax=\lambda x$
    \State set label eigenvector
    \EndFor
\State \textbf{End}
\end{algorithmic}
\label{algo:eigenvector}
\end{algorithm}
\end{enumerate}

\subsubsection{Procedure Go}
To validate our model, we apply routing techniques on our developed model. There are two routing approaches that we used in our work. The first one is random walk and the second one is centrality-based routing. Next, we present the details and specification models of these two routing approaches described as procedures. 

\begin{enumerate}
\item Procedure random-walk:\\
In case of a random walk, the walkers are set initially on the source nodes. They search their neighbors and select one of them randomly. This process goes repeatedly until all the target nodes have been visited.
Let us consider an undirected graph $G(V,E)$ , a random walk is a stochastic process that starts from a given vertex, then select one of its neighbors randomly to visit next. It has no memory that keeps information of previous moves. It stops when the termination condition meets.
\begin{algorithm}[H]
\caption{Procedure \textbf{random-walk:} Used for routing on network}
\emph{Input:} All input parameters provided by user interface\\
\emph{Output:} End to end delivery from sources to the destinations\\
\emph{Execution:} Called repeatedly on simulation execution\\
\textbf{begin}
\begin{algorithmic}[1]
 \State start from sources
 \While {$is-finish \not= true$}
 \State search neighbors
 \State select any of neighbor
 \State \textbf{move} to the selected
\EndWhile
\end{algorithmic}
\textbf{End}
\label{algo:random-walk}
\end{algorithm}

\item Procedure centrality-rw:\\
The procedure ``centrality-rw'' is another approach which is used in our work for routing purpose. The working of this technique is as follows. The Walker search for neighbor nodes. They take a decision on the basis of centrality. They search for that neighbor node which has maximum centrality value. Then move to the selected node. If they found no node with maximum centrality value, then they select randomly a node from neighbors. The working of centrality-base routing is shown in figure \ref{fig:fcCR}.
\begin{figure}[H]
\begin{center}
\includegraphics[width=\linewidth, height=8.0 cm]{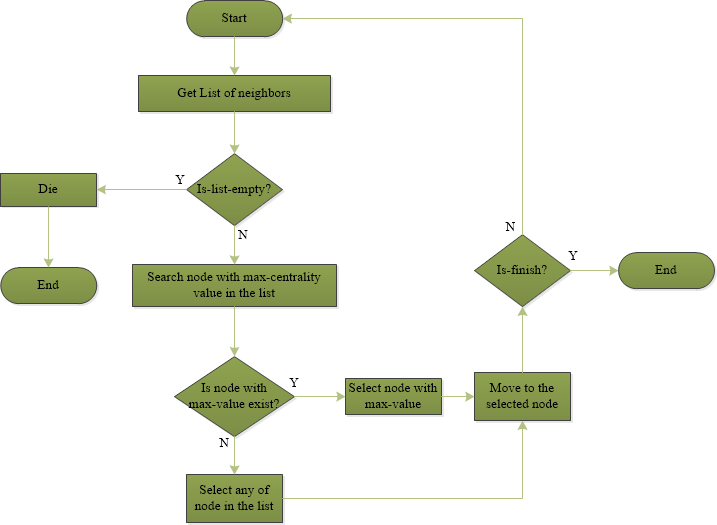}
\caption{Flowchart for centrality-based routing algorithm. We can see that the algorithm starts with making a list of neighbor nodes of the current location. In the next step, it checks the content of the list. If the list is empty, then the simulation stops, in case no, the algorithm search for the node having maximum centrality value. In case, yes, it selects that node, in case no, it selects any node from the neighbor list. The next step is to move to the selected node. After this, it checks the finish goal, in case yes, the simulation stops and in case no the control goes to the first step.}\label{fig:fcCR}
\end{center}
\end{figure} 
\begin{algorithm}[H]
\caption{Procedure \textbf{centrality-rw:} Used for routing on network}
\emph{Input:}All input parameters provided by user interface\\
\emph{Output:} End to end delivery from sources to the destinations\\
\emph{Execution:} Called repeatedly on simulation execution\\
\textbf{begin}
\begin{algorithmic}[1]
 \State start from sources
 \While {$is$-$finish \not= true$}
 \State get-list-of-neighbors
  \If {is-list-empty=true}
 	\State die
    \State stop
    \EndIf
  \State Else
  		\State Search node with max-centrality in the list
        \If {node with max-value exist= true}
        	\State select node with max-value
         \EndIf
         \State Else
         \State select any of node in the list
 \State \textbf{move} to the selected
\EndWhile
\end{algorithmic}
\textbf{End}
\label{algo:centrality-rw}
\end{algorithm}

\item Generate plot:\\
``do-plot'' procedure is used to plot on the execution of the simulation of the model. Here, this procedure plots two types of information. The first, at each time step, it counts the number of nodes that have been visited. The second, it counts the number of consumers that are visited. Next, we present model specification and detail for the experiment in our model.

\begin{algorithm}[H]
\caption{Procedure \textbf{generate-graph:} Generate graphs of the current simulation state}
\begin{algorithmic}
\State \emph{Input:} No input required
\State \emph{Output:} Generate plot on simulation execution
\State \emph{Execution:} Called by Go procedure
\State \textbf{begin}
	\State Plot the number of nodes that are visited
    \State Plot the number of targets visited
\State \textbf{End}
\end{algorithmic}
\label{algo:do-plot}
\end{algorithm}

\end{enumerate}
\subsubsection{Performed experiment}
Two types of experiments are performed in our model simulation. The first one is used for random walk algorithm; the second is used for centrality-based routing algorithm. 
\begin{algorithm}[H]
\caption{Experiment: Demonstrates the effect of random walk and centrality routing on different networks}
\begin{algorithmic}
\State \emph{Input:} $< num-nodes, num-source, num-target, num-walker, network-type>$
\State \emph{Setup Procedure:} $<setup>$
\State \emph{Go Procedure:} $<random-walk, centrality-rw>$
\State \emph{Repetition:} 10\\
\hrulefill
\State \emph{Input:}
	\State \hspace{\algorithmicindent} \textbf{num-nodes:} 500
    \State \hspace{\algorithmicindent} \textbf{num-source:} 5, 10, 50, 100
    \State \hspace{\algorithmicindent} \textbf{num-targets:} 50, 100, 150, 200
    \State \hspace{\algorithmicindent} \textbf{num-walker:} 5, 10, 50, 100
    \State \textbf{networ-type:} small-world, scale-free, random-network
\State \emph{Reporter:}
	\State \hspace{\algorithmicindent} \textbf{measure-nodes:} count number of nodes that have been visited
	\State \hspace{\algorithmicindent} \textbf{measure-targets:} count number of targets that have been visited
\State \emph{Stop condition:}
	 \State \hspace{\algorithmicindent} If all targets are visited, Stop
\end{algorithmic}
\end{algorithm}
\label{algo:expt}

\subsection{Centrality-based routing algorithm: Time complexity analysis}
In this section, we present the centrality-based routing algorithm time complexity analysis. The time complexity of our proposed centrality-based routing is a linear function of $n$ that is $\mathcal{O}(n)$. In algorithm analysis, we analyze the cost and number of times that each step takes for execution.  All steps take a constant time, except step 2 and 8 which takes n time for execution. In step 2,  a while loop executes for n times. In step 8, algorithm searches in the list item of nodes and then selects the node with the largest value, so it takes $n$ time. The total running time is the sum of running time and cost of each step in the algorithm.
\begin{algorithm}[H]
\caption*{Procedure \textbf{centrality-rw:} Used for routing on network}
\emph{Input:}All input parameters provided by user interface\\
\emph{Output:} End to end delivery from sources to the destinations\\
\emph{Execution:} Called repeatedly on simulation execution\\
\textbf{begin} \hfill Cost \hspace{4ex} time
\begin{algorithmic}[1]
 \State start from sources \hfill $C_1$ \hspace{4ex}1
 \While {$is-finish \not= true$}\hfill $C_2$ \hspace{4ex}$n$
 \State get-list-of-neighbors \hfill $C_3$ \hspace{4ex}1
  \If {is-list-empty=true} \hfill $C_4$ \hspace{4ex}1
 	\State die \hfill $C_5$ \hspace{4ex}1
    \State stop \hfill $C_6$ \hspace{4ex}1
    \EndIf 
  \State Else
  		\State Search node with max-centrality in the list \hfill $C_7$ \hspace{4ex}$n$
        \If {node with max-value exist= true} \hfill $C_8$ \hspace{4ex}1
        	\State select node with max-value \hfill $C_9$ \hspace{4ex}1
         \EndIf
         \State Else
         \State select any of node in the list \hfill $C_10$ \hspace{4ex}1
 \State \textbf{move} to the selected \hfill $C_11$ \hspace{4ex}1
\EndWhile
\end{algorithmic}
\textbf{End}
\label{algo:cmp-cr}
\end{algorithm}
\begin{equation}\label{eq:algocomplexity}
T(n)=C_1 (1)+C_2 (n)+C_3 (1)+C_4 (1)+C_5 (1)+C_6 (1)+C_7 (n)+C_8 (1)+C_9 (1)+C_10 (1)+C_11 (1)
\end{equation}

\begin{equation}\label{eq:algcmp2}
 c =(C_1+C_3+ C_4+C_5+C_6+C_8+C_9+C_10+C_11) (1)
\end{equation}
\begin{equation}\label{eq:algcmp3}
C_2(n)+C_7(n)=(C_2+C_7)n
\end{equation}
\begin{equation}\label{eq:algcmp4}
 a=C_2+C_7
\end{equation}
By putting Eq.\ref{eq:algcmp2} and Eq.\ref{eq:algcmp4} values in Eq.\ref{eq:algocomplexity}, we get:
\begin{equation}\label{eq:finalcmp}
T(n)=a(n)+c
\end{equation}
\begin{equation}
T(n)=\mathcal{O}(n)
\end{equation}
\section{Results and Discussion}\label{sec:resultanddiscusion}
In this section, we present results obtained from DREAM methodology, then we compare ODD vs DREAM followed by empirical analysis and related work.
\subsection{Complex network of the model}\label{sec:cnmodel}
\begin{figure}[H]
  \includegraphics[width=\linewidth]{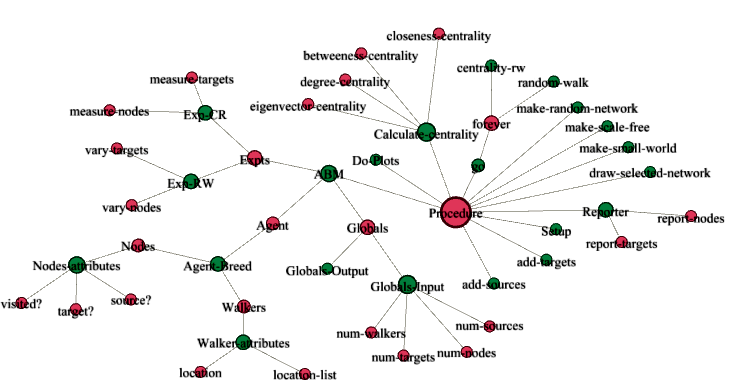}
  \caption{A network model of our proposed ABM for the smart grid. We can see that in the network the root node is ``ABM'' which is connected to the four nodes named as ``globals, agents, procedure, and expts''. The ``globals'' node is connected to the globals output and input parameters. The ``agent'' node is connected with the involved entities in the model. The ``procedure'' node is connected with all other processes and functions in the model. The ``expts'' node is connected with all experiments carried out in the model simulation.}
  \label{fig:nw-final}
\end{figure}

Figure \ref{fig:nw-final} shows the complex network of our proposed ABM. This network presents our ABM in a visualized form. We develop the network model of our proposed ABM using Gephi (a network toolkit).

First, we start from the root node ``ABM''. This node is expanded into leaf nodes ``global variables, agent, procedures, and expts''.  The global variables are further expanded into global output and input. These are input provided by the user interface to the model. ``Agent'' node represents the involved entity in the model which is further expanded into two types of agents that are named as node and walker.  After defining global inputs, agents, our next focus is on ``procedure'' node. The ``procedure'' node is the root for all procedures used in the model. This node has the highest node degree in the network model.  The node ``expts'' is a parent for all sub-nodes that represent different experiments carried out in model simulation. 

\subsection{Social network analysis}\label{sec:sna}
In this section, we present the results obtained by applying social network analysis(SNA) tool on the network model. The SNA provides quantitative measures to give network topological details. Using these quantitative measures we can perform a comparison of different network models. 

\begin{figure}[H]
  \includegraphics[width=\linewidth]{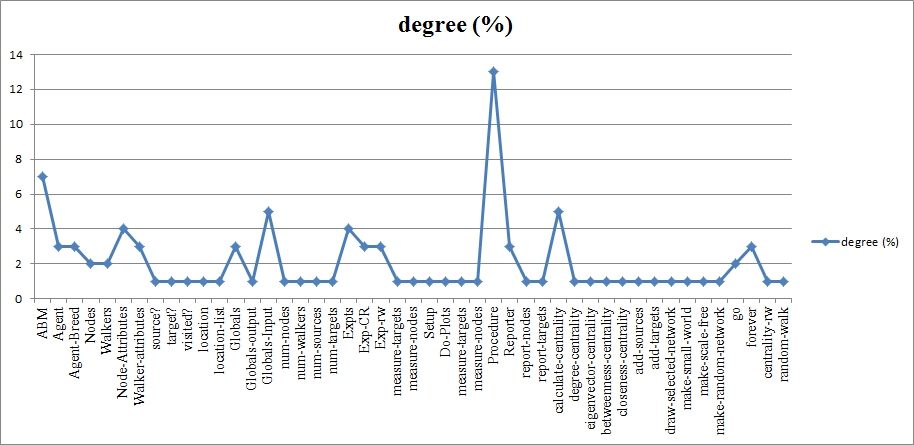}
  \caption{Degree centrality of the network: It shows \emph{Prodecdure} node has the highest degree in the network.}
  \label{fig:degree-centrality}
\end{figure}

In figure \ref{fig:degree-centrality} degree centrality of the network is plotted. It shows that the procedure node has highest centrality. Next the ABM node and third is global-input.

\begin{figure}[H]
  \includegraphics[width=\linewidth]{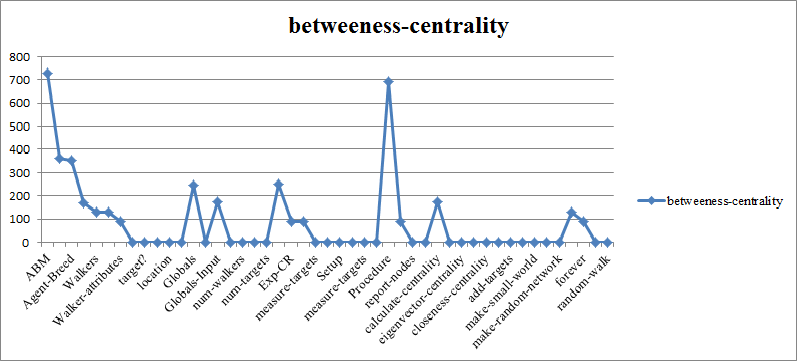}
  \caption{Betweenness centrality of the network: \emph{ABM} node has the highest betweeness centrality value in the network. }
  \label{fig:betweenness-centrality}
\end{figure}

In figure \ref{fig:betweenness-centrality} betweenness centrality is presented. It demonstrates that ABM node has highest betweenness centrality and the second highest betweennes centrality demonstrated by procedure node.

\begin{figure}[H]
  \includegraphics[width=\linewidth]{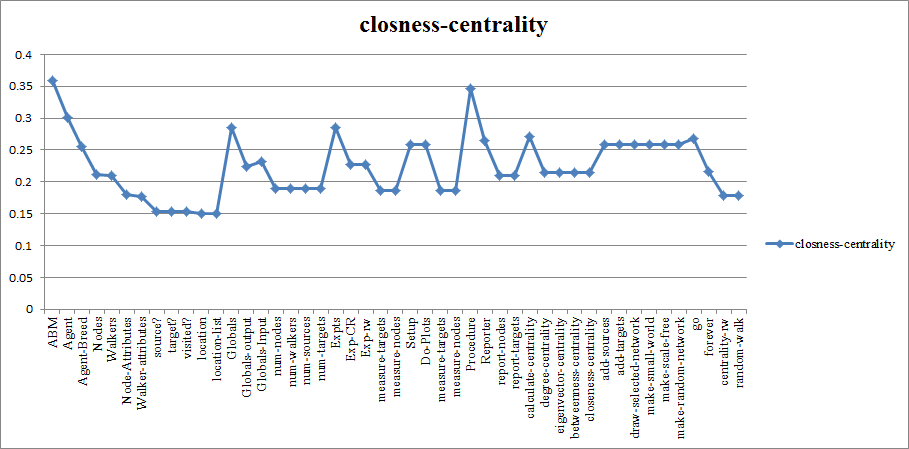}
  \caption{Closeness centrality of the network: \emph{ABM and Procedure} node has the highest closeness centrality value.}
  \label{fig:closeness-centrality}
\end{figure}

Figure \ref{fig:closeness-centrality} shows closeness centrality of the network. It demonstrates that ABM node has highest closeness centrality and procedure node is on second number.

\begin{figure}[H]
  \includegraphics[width=\linewidth]{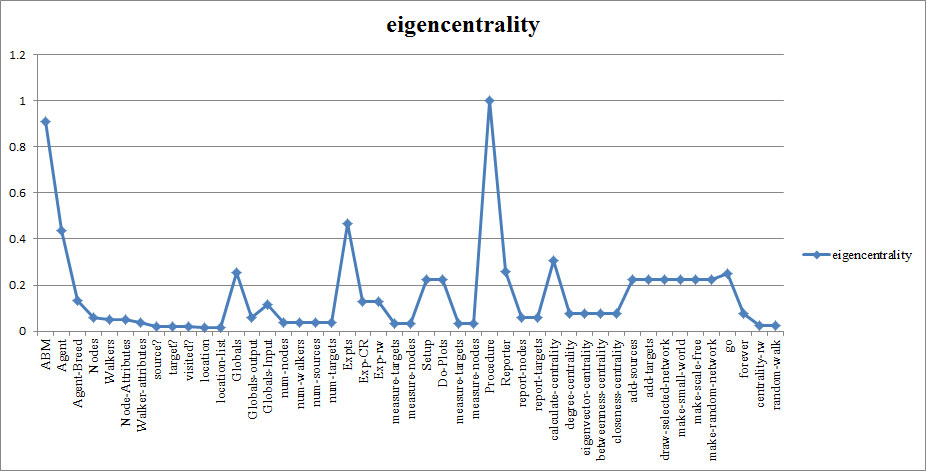}
  \caption{Eigenvector centrality of the network: It shows \emph{Procedure} node is on the top list in the network model.}
  \label{fig:eigenvector-centrality}
\end{figure}

Figure \ref{fig:eigenvector-centrality} shows eigenvector centrality of the network. It shows that procedure node has highest eigenvector centrality followed by ABM node.

\subsection{Comparison of ODD and DREAM}\label{sec:odddream}
In this section, we provide a qualitative as well as quantitative comparison between ODD and DREAM specification techniques. 

The ODD specification allows a textual based description of ABM with the purpose to make model readable and promotes the rigorous formulation of models. It comprises checklist that covers key features through which one can describe an ABM. The ODD specification has some limitations that are described in the following.

The ODD specification only provides a textual based description of ABM. Sometimes for large ABM, it has less description, which is insufficient to cover all the features of the ABM. It has no quantitative assessment of the ABM on the base of which one can perform a comparison between different ABM. Reviewing and comparison of different ABM is difficult. For comparison and classification purpose, the only possible way is to make a table and put ODD checklists of different ABM. Then search for similarities and differences.

According to the Grimm et al. \cite{grimm2010odd}, a survey was conducted from 2006 to 2009 of those publications in which ODD was used. According to this survey, only $75 \%$ publications used ODD correctly, while $25 \%$ publications used ODD incorrectly and some parts of the protocol were compromised. Grimm et al. \cite{grimm2010odd} formulates the conclusion that it is difficult to write an ABM specification by following ODD protocol.

Another issue which is identified in the ODD specification is redundancy. Some parts of the specification like purpose section is also included in the introduction section of the document. the design concept section is also repeated in the sub model section of the method. Sub-model section is repeated in the process and scheduling section.

Sometimes there may be different publications with a different version of the same ABM. Then these publications have the same ODD with little modification in entities and process section. Another limitation of ODD is that the textual based description is too specific which is not useful for replication of ABM; there exist some ambiguities and misunderstanding about ABM.

On the other side, DREAM allows a detailed specification of ABM. It comprises of making a complex network of the model, pseudo-code specification, and network analysis steps. 

This method allows for inter-disciplinary comparative study and communication among different scientific domains. So, if a model is developed in a social science, it can be compared visually and quantitatively with a model developed in biological science and vice versa. For an instance, a social model for aid spreading presented in \cite{niazi2013modeling} and biological model developed for the emergence of snake-like structure in \cite{niazi2014emergence}. By developing complex networks of both ABM models, we can easily compare and analyze both networks in the same manner. DREAM specification approach can be applied to any ABM of any research domain.

DREAM provides to present ABM in the complex network model. This allows reading and understanding ABM visually without going to the exact code specification. By performing network analysis on this network of ABM, it also gives a quantitative measurement of the ABM. These quantitative measures are the digital footprint of ABM and can be used to compare different ABM.

DREAM further allows pseudo-code specification and details of the ABM. This specification helps in understanding ABM completely without regards discipline. This specification then offers a translation to the code and developing ABM. It concludes that by using DREAM, any ABM can be replicated easily.

Next, we present an empirical analysis of the ODD and DREAM methodologies. We evaluated these methodologies based on 13 features as shown in table \ref{tbl:features}. We used H=2, M=1 and L=0 for evaluation purpose that demonstrates which methodology offers which feature to what extent. Then we computed the rank of each methodology by averaging results shown in table \ref{tbl:analysisODDDREAM}.  

\begin{table}[H]
\begin{center}
\caption{Selected features for empirical analysis of ODD and DREAM}\label{tbl:features}
\begin{tabular}{ p{3.5cm} | p{9.0cm}}

\hline
\textbf{Feature} & \textbf{Description} \\ \hline
Social and technical process & Considers the level of social and technical aspects in the methodology \\ \hline
Adaptability & Referring how much the methodology is flexible to be adopted in different domains \\ \hline
Reusability & It refers the extent of the methodology to be used for model replication \\ \hline
Redundancy & Concerns with the number of repeated modules in the methodology \\ \hline
Stepwise & Measures how much methodology is based on sequential steps \\ \hline
Documentation & It involves the process of documenting the model  \\ \hline
Network model & It is the developing of network for the model \\ \hline
Pseudo-code & Concerns with the presenting Pseudo-code specification for the model \\ \hline
Network analysis  & Concerns to applying social network analysis tool on network \\ \hline
Transition & Refers the extent of conversion to model development \\ \hline
Communication & Refers to the level of communication among multiple disciplines  \\ \hline
Confusion & It shows the degree of complexity in the methodology \\ \hline
User satisfaction & Refers to the level of convenience\\ \hline
\end{tabular}
\end{center}
\end{table}

\begin{table}[H]
\begin{center}
\caption{Empirical analysis of ODD vs DREAM}\label{tbl:analysisODDDREAM}
\begin{tabular}{l | l | l}
\textbf{Feature} & \textbf{ODD} & \textbf{DREAM}\\  \hline
F(1) & H & H \\ \hline
F(2) & L & H \\ \hline
F(3) & L & H \\ \hline
F(4) & M & L \\ \hline
F(5) & H & H \\ \hline
F(6) & H & M \\ \hline
F(7) & H & H \\ \hline
F(8) & L & H \\ \hline
F(9) & L & H \\ \hline
F(10) & L & H \\ \hline
F(11) & L & H \\ \hline
F(12) & L & H \\ \hline
F(13) & L & H \\ \hline
Rank & $ 0.69 $ & $ 1.76 $ \\ \hline
\end{tabular}
\end{center}
\end{table}
\subsection{Simulation results}
\subsubsection{Simulation setup}
To simulate a smart grid based complex scenario, we developed small-world, scale-free and random complex network using Agent-Based Modeling approach. In order to validate our work, we applied routing techniques such as random walk and centrality-based routing on large-scale complex networks, specifically in the smart grid domain. For comparison, we applied random and centrality-based routing on these networks and analyzed their behavior on these networks.
\subsubsection{Evaluation metrics}
For performance evaluation purpose, we used average delivery rate parameter. The average delivery rate is defined as the number of packets sent by sources and successfully received by consumers. Mathematically can be written as following:
\begin{equation}\label{eq:avgrate}
\sum_{1}^{n}\frac{Ds}{Dc} 100
\end{equation}
Where Ds represents data packets sent by the source and Dc represents data packets received by consumers. The experiment was performed for different case studies such as a different number of consumers and generation units. Then the simulation results were averaged over 30 executions.
To see the behavior of routing techniques, when going from source locations towards the destinations through different paths at each time step. We used different parameters and observed for which combination it takes less convergence time. The simulation environment is set according to the parameters as shown in table \ref{tbl:metrics}.
\subsubsection{Results of Random walk}
We applied random walk routing technique on different complex networks. The simulation results demonstrate that random walk routing technique showed less iteration in the case of small-world topology as compared to others network topologies.

\begin{figure}[H]
\begin{subfigure}{.5\textwidth}
  \centering
  \includegraphics[width=1.0\linewidth]{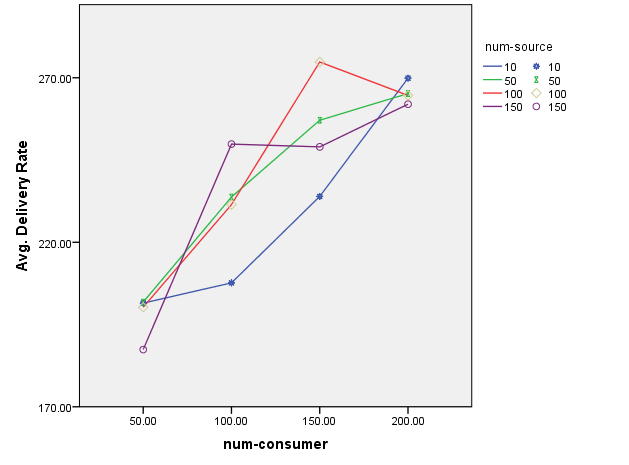}
  \caption{Random Walk on Small-world network}
  \label{fig:rwsm}
\end{subfigure}%
\begin{subfigure}{.5\textwidth}
  \centering
  \includegraphics[width=1.0\linewidth]{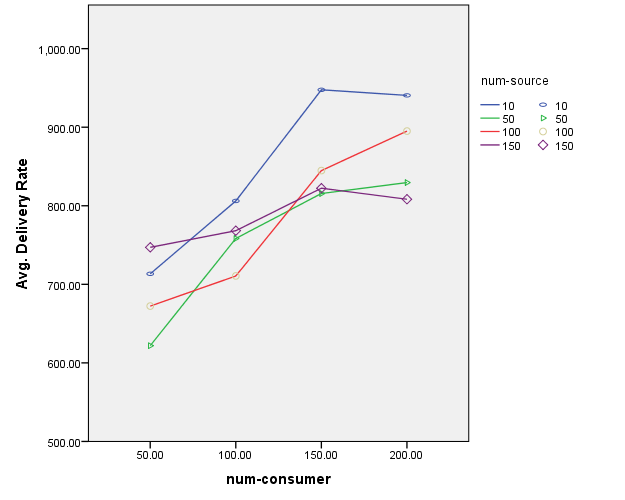}
  \caption{Random Walk on scale-free network}
  \label{fig:rwsf}
\end{subfigure}
\caption{The simulation results of Random walk on different networks: Part(a) shows results of a random walk on the small-world network. The x-axis shows a number of consumers and the y-axis shows average end to end delivery rate against a different number of sources. Part(b) shows results of random walk for different number of consumers and sources on scale-free network of five hundred nodes.}
\label{fig:rwswsf}
\end{figure}

For small world network, figure \ref{fig:rwsm} shows the simulation result for different number of sources and consumers. This shows convergence rate on different case studies. The results show convergence rate lies between 180-280 iterations. 

For scale-free network, figure \ref{fig:rwsf} shows simulation results for different numbers of sources and consumers. Random walk shows large convergence rate as compared to small world network.

\begin{figure}[H]
\begin{subfigure}{.5\textwidth}
  \centering
  \includegraphics[width=.8\linewidth]{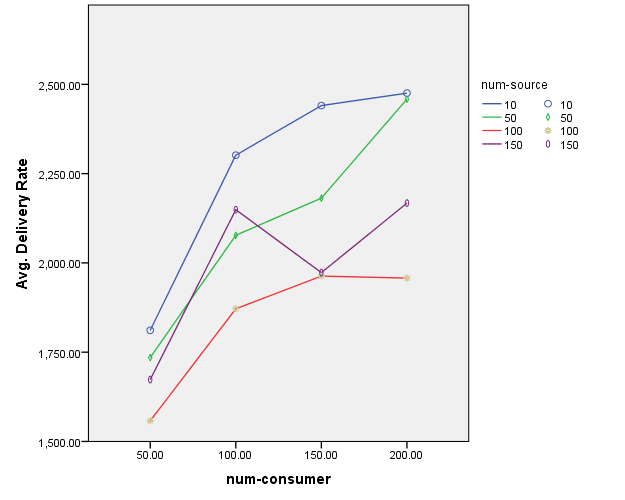}
  \caption{Random Walk on random network}
  \label{fig:rwrn}
\end{subfigure}%
\begin{subfigure}{.5\textwidth}
  \centering
  \includegraphics[width=.8\linewidth]{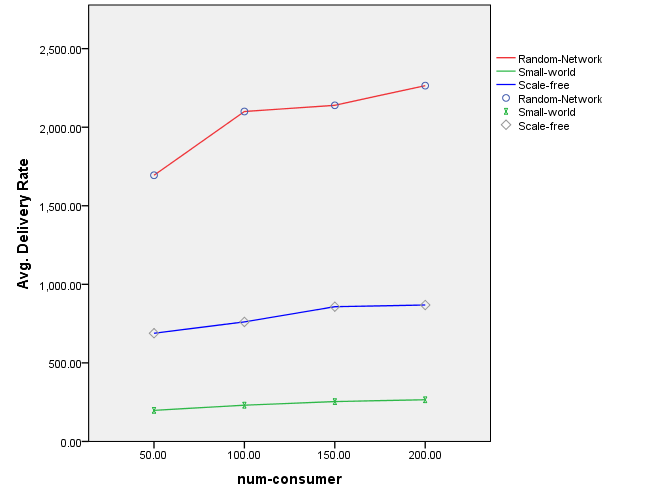}
  \caption{Random Walk on overall networks}
  \label{fig:rwo}
\end{subfigure}
\caption{The simulation results of random walk: Part(a) shows random walk results on a random network with five hundred nodes. Part(b) shows random walk results on overall complex networks. Results of all three networks are compared with(number of consumers: 50, 100, 150, 200, number of sources: 10, 50, 100, 150).}
\label{fig:rwrno}
\end{figure}

For random network, figure \ref{fig:rwrn} shows simulation results for different numbers of sources and consumers. In this network topology, random walk demonstrate very large convergence rate as compared to both scale-free and small world. This is due to the network topology.

Figure \ref{fig:rwo} shows the performance of random walk on different network topologies. The small-world topology demonstrates less iteration while random network has very large convergence rate.

\subsubsection{Issues identified in Random walk}
\begin{enumerate}
\item Agents can move randomly on the network, they select a random node from their neighbor list.
\item Agents can move to previously visited nodes.
\item Agents do not maintain records when traversing nodes of the network.
\item Sometimes, agents get stuck on the network, this increases computational time.
\end{enumerate}

\subsubsection{Results of Centrality-based routing}
\begin{figure}[H]
\begin{subfigure}{.5\textwidth}
  \centering
  \includegraphics[width=.8\linewidth]{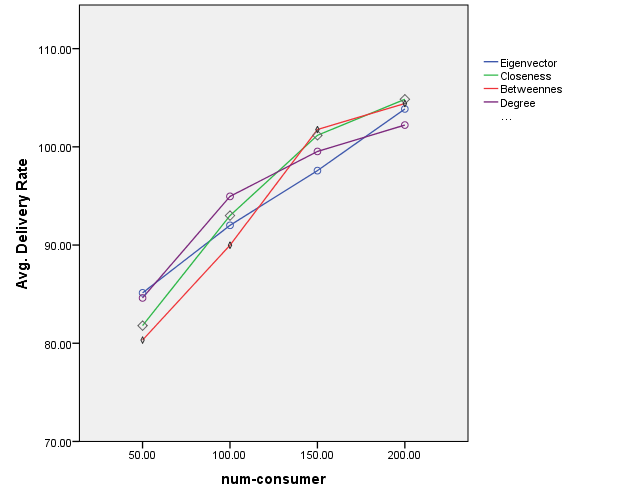}
  \caption{Centrality-based routing on small-world network}
  \label{fig:crsw}
\end{subfigure}%
\begin{subfigure}{.5\textwidth}
  \centering
  \includegraphics[width=.8\linewidth]{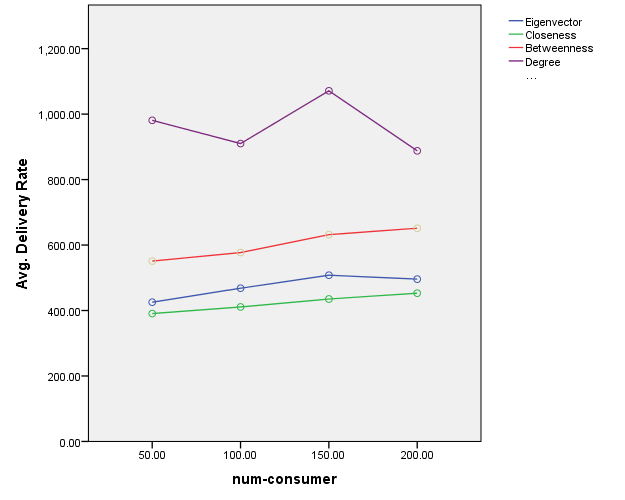}
  \caption{Centrality-based routing on scale-free network}
  \label{fig:crsf}
\end{subfigure}
\caption{The simulation results of Centrality-based routing on complex networks. Results for different types of centrality routing based on Closeness, Betweenness, Eigenvector, and Degree are plotted. The experiments were performed for consumers: 50, 100, 150, 200 and sources: 10, 50, 100, 150: Part(a) shows centrality-based routing on the small-world network. Part(b) shows centrality-based routing results on scale-free network.}
\label{fig:crswsf}
\end{figure}

Figure \ref{fig:crsw} shows centrality routing(CR) on small-world network with different numbers of consumers and generating units. The simulation results show that on average, each CR has equal convergence rate. When it is compared with other complex networks, it is found through simulation results that CR on small-world has less convergence rate compared with other networks. In this case, CR on small-world has convergence rate between 80 to 105 iterations.

Figure \ref{fig:crsf} shows simulation results of the CR on scale-free network with different numbers of consumers and generating units. The simulation results demonstrate that on scale-free network, Degree-CR has high convergence rate as compared to other approaches.

\begin{figure}[H]
\begin{subfigure}{.5\textwidth}
  \centering
  \includegraphics[width=.8\linewidth]{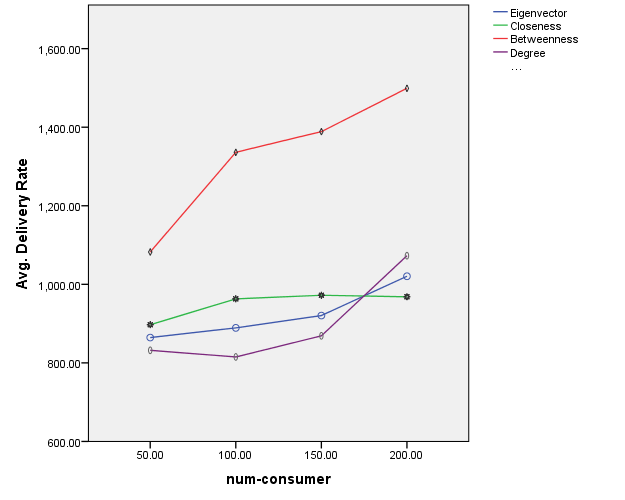}
  \caption{Centrality-based routing on random network}
  \label{fig:crrn}
\end{subfigure}%
\begin{subfigure}{.5\textwidth}
  \centering
  \includegraphics[width=.8\linewidth]{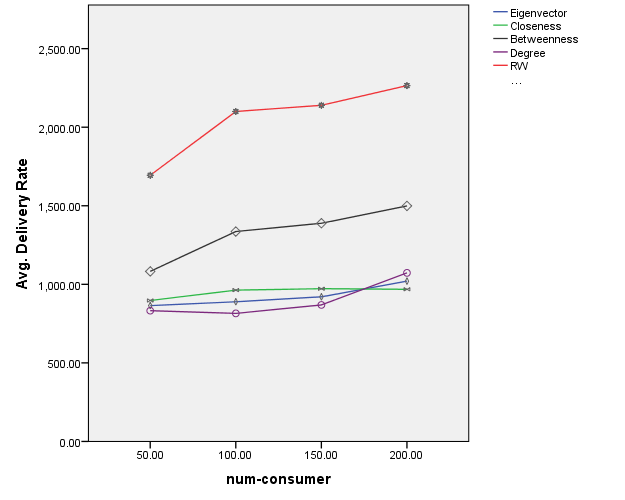}
  \caption{RW vs CR on random network}
  \label{fig:rwcrrn}
\end{subfigure}
\caption{Part(a) shows simulation results of different centrality-based routing algorithm on a random complex network comprises of five hundred nodes. The number of consumers: 50,100,150,200 and number of sources: 10,50,100 are used. In part(b), the simulation results of random-walk and centrality-based routing algorithms are compared for different number consumers and sources.}
\label{fig:crrnrwcrrn}
\end{figure}

Figure \ref{fig:crrn} shows simulation results of Degree-CR, Closeness-CR, Betweenness-CR and Eigenvector-CR applied on random network using different numbers of consumers and generating unit. Then the obtained results were plotted according to the figure \ref{fig:crrn}. This shows that Degree-CR has less convergence time as compared to others centrality routing techniques. 

\subsubsection{Random-walk vs Centrality-based routing}
Figure \ref{fig:rwcrrn} shows simulation results of different routing techniques on random network. It demonstrates that random walk has large iteration as compared to other routing techniques.

\begin{figure}[H]
\begin{subfigure}{.5\textwidth}
  \centering
  \includegraphics[width=.8\linewidth]{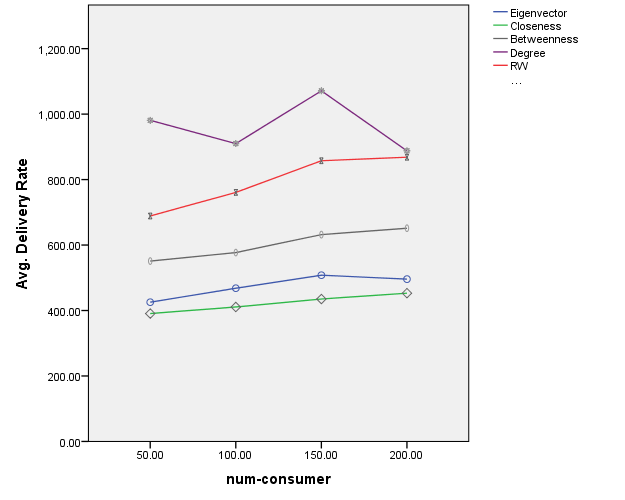}
  \caption{RW vs CR on scale-free network}
  \label{fig:rwcrsf}
\end{subfigure}%
\begin{subfigure}{.5\textwidth}
  \centering
  \includegraphics[width=.8\linewidth]{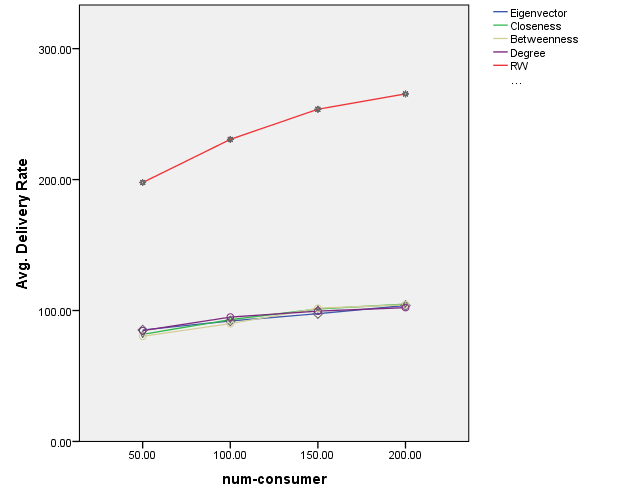}
  \caption{RW vs CR on small-world network}
  \label{fig:rwcrsw}
\end{subfigure}
\caption{Comparison of random-walk and centrality-based routing algorithms on different networks. Experiments were performed for consumers:50,100,150,200 and source:50,100,150,200. Part(a) shows comparative results of random-walk and centrality-based routing on scale-free network. Part(b) shows comparative results of random-walk and centrality-based routing on small-world network. The simulation results of centrality routing based on closeness, betweenness, eigenvector and degree are same for small-world network.}
\label{fig:rwcrsfrwcrsw}
\end{figure}

Figure \ref{fig:rwcrsf} shows simulation results of different routing techniques on scale-free network. It demonstrates that Degree-CR and random walk have large convergence time.

Figure \ref{fig:rwcrsw} shows the simulation results of different routing techniques on small-world network. It demonstrates that CR techniques have a similar convergence rate while random walk has large convergence rate. 

The total time reduction of centrality-based routing compared with the random-walk algorithm on the random network, scale-free network, and small-world network are shown in tables \ref{tbl:ECRN},\ref{tbl:ECSF},\ref{tbl:ECSW} respectively.

\subsection{Comparison with previous work}\label{sec-relatedwork}
In this section, we present some related work. Previously, the agent-based and complex network-based approaches have successfully been used in the smart grid. However, there is no such model with the conjunction of specification aspect of the model. Next, we present some of the previous studies from an agent-based and complex network perspective.

In \cite{hinker2017novel}, a conceptual model is developed for energy domain. This model is integrated with the ODD methodology for documenting ABM. In this model, some other concept was added like layers, objects, actor and working point to bridge between the social and technical system in the energy domain. However, this conceptual model was not validated on ABM. 

In \cite{jiang2017check},  a check-in based routing approach is proposed for network traffic model. In this work, betweenness centrality used to assign node as the check-in node between source and destination. The proposed routing strategy was implemented on the scale-free network. However, the optimization of routing remained as the open problem of this work. 

In \cite{niazi2009agent}, Niazi and Hussain have presented agent-based tools for modeling and simulation of self-organizing in wireless sensor network. They demonstrated the usability of agent-based tool NetLogo, and developed different experiments that show how to model different scenario in the sensor network domain.

Batool and Niazi in \cite{batool2017modeling}, presented a novel hybrid approach for modeling internet of things (IoT)-based complex scenario using standard complex networks such as small-world, scale-free, random and lattice network.
In \cite{batool2014towards}, novel agent-based and complex network-based modeling approaches are used in the wireless sensor network domain. In this work, different standard complex network such as small-world, scale-free, and random network is used. Further, random walk routing strategy is implemented for communication between different sensors in the network.
 
In \cite{pradittasnee2017efficient}, another routing technique proposed for large-scale sensor network-based environment. In this work, local and global update strategy is introduced for maintenance and efficient routing in the network. This approach monitors any changes in network and update routing path according to the situation. Results demonstrate the effectiveness of the techniques and reduced end to end delivery rate as compared to the previous techniques. 

In \cite{wang2017synchronisation}, research work is carried out on frequency synchronization in the power grid system. In this work, the network theory concept was used to monitor, control and exploit the frequency variation of the power system.

In \cite{jia2012security}, research work is carried out on security analysis using complex network approach in power system. In this work, the power adjacency matrix approach is proposed for analysis and measurement of the power flow and activities of each node and links on the network.

In \cite{guan2011routing}, proposed a novel routing strategy based on betweenness centrality in a complex network. In this work scale-free network is used and routing was performed based on expanding betweenness of each node. This method shifts the load from the node with higher betweenness to the lower. 

Liu et al. in \cite{yan2017efficient}, proposed an efficient probability routing strategy using scale-free complex network topology. This method utilizes the probability concept for redistributing load from critical nodes to the non-critical nodes. Results showed that routing path is reduced $30 \%$ as compared with previously probability routing technique. 

In \cite{oliveira2010centrality}, research work is carried out on routing strategy in wireless sensor network and proposed sink betweenness distributed routing algorithm. In this approach, betweenness of each node is calculated in which sink node exists as a terminal node. This work was implemented on the random network.

Ansari et al. in  \cite{ansari2016multi}, proposed a multi-agent system for reactive power control system in a smart grid. This work reduced power losses and provided exploitation of available power resources. Zhang et al. in \cite{zhang2016agent}, presented a voltage variation control strategy. This approach controls voltage profile in the specified range of the studied system, which results in reducing system loss and improving system reliability. Authors in \cite{eriksson2015multiagent, ghorbani2016multiagent} have worked on fault location and restoration in smart grid by using the complex network approach. These studies demonstrate the modeling of fault location and restoration process in a distributed power network. Weng et al. in \cite{ weng2017fault} also worked on fault location and proposed the use of particle swarm optimization technique for locating voltage disturbance sources in a distributed power grid. 

Regarding communication management, different studies also have been presented in smart grid. Wang et al. in \cite{wang2013adaptive}, proposed an adaptive strategy for energy trading between utility grid and consumers. In this work, each agent can communicate with each other for sharing information about energy usage and cost. Tsai et al. in\cite {tsai2017communication} have worked on distributed large-scale consumers load with the conjunction of renewable energy resources. In this work, a neighbor communication strategy is applied. This results in low communication cost.  E. Kremers et al. in \cite{kremers2013multi} presented a bottom-up approach for the smart grid modeling. It consists of two layers; physical layer for electrical power transmission and logical layer for communication. This model has the ability to integrate new devices in the smart grid environment. It provides dynamic load management, power, communication control and monitoring.
 
Regarding power scheduling, In \cite{Chao2016}, Chao and Hsiung proposed Fair Energy Resource Allocation (FERA) algorithm for electricity trading in smart grid. This technique prevents starvation situation and fatal problem, it also reduces power cost.
In  \cite{li2015consensus} Li et al.  proposed a look-ahead scheduling model for flexible loads in a smart grid environment. This model provides flexible strategies to handle the large flexible load. In \cite{wang2016reinforcement} have proposed a multiple microgrid based model, in which each micro grid trades energy to the market and randomly selects their strategy for maximizing their revenue.  Samadi et al. in \cite{samadi2016load} worked on load scheduling and power trading with a combination of large scale renewable energy resources. This model allows users to consume local energy. 
In \cite{clausen2017agent} an agent-based model is proposed for large scale virtual power plant. The proposed model constitutes of complex, heterogeneous distributed energy resources with a combination of local and multi-objective function.
In \cite{shirzeh2015balancing}, Shirzeh et al. worked on management of renewable energy resources and storage devices in a smart grid environment. They proposed a multi-agent system based on a plug-and-play technique for management and controlling storage resources in the smart grid. In \cite{kuznetsova2013reinforcement}, Kuznetsova et al. presented two step-ahead reinforcement learning algorithm for battery scheduling within microgrid architecture. It is composed of local consumers, generator and storage devices connected to the external grid. This technique predicts and forecasts power demand of consumers.
\section{Conclusions}\label{sec:conclusion}
In this paper, we proposed modeling and simulation of the smart grid by using a combination of agent-based and complex network-based approaches. To further validate our work, we applied routing techniques such as random walk and centrality-based routing. We used different standard complex networks such as small-world, scale-free and random network. The simulation results demonstrate that centrality-based routing gives better results on the small-world network as compared to other networks. For ABM specification, we followed two approaches, the one is ODD and the other is DREAM methodology. We presented qualitative as well as a quantitative comparison of both ODD and DREAM specification techniques. The comparative study of ODD and DREAM proved that DREAM methodology is the more useful approach for documenting an ABM not only in terms of modeling but also for replication of the models, specifically in the smart grid domain.
\newpage
\begin{figure}[H]
\begin{center}
\includegraphics[width=\linewidth, height=9.0 cm]{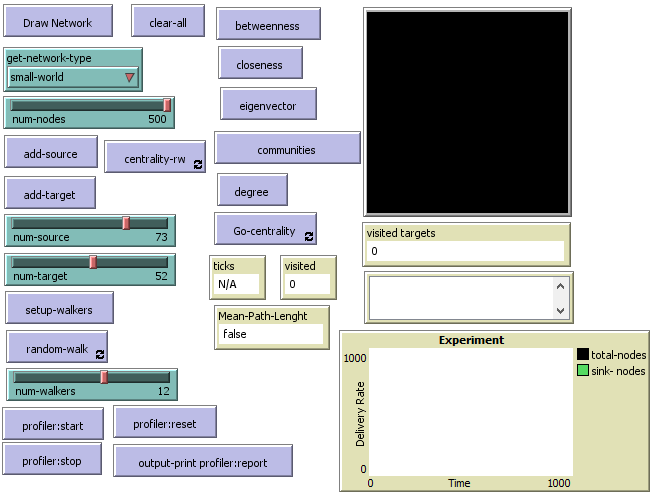}
\caption{Screenshot of the developed ABM of smart grid. The image shows user interface of NetLogo simulation tool. It consists of sliders, chooser, monitors, buttons, and world ( a simulation environment).}
\label{fig:main-abm}
\end{center}
\end{figure}

\begin{figure}[H]
\begin{subfigure}{.5\textwidth}
  \centering
  \includegraphics[width=.9\linewidth]{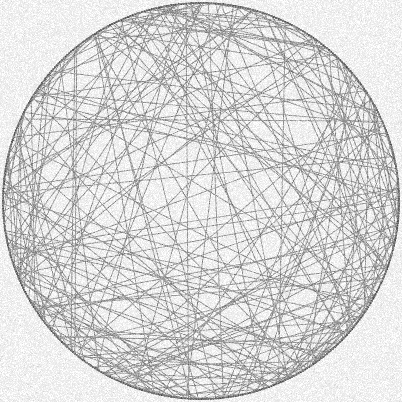}
 \caption{Small-world network }\label{fig:sw-nw}
\end{subfigure}%
\begin{subfigure}{.5\textwidth}
  \centering
  \includegraphics[width=.9\linewidth]{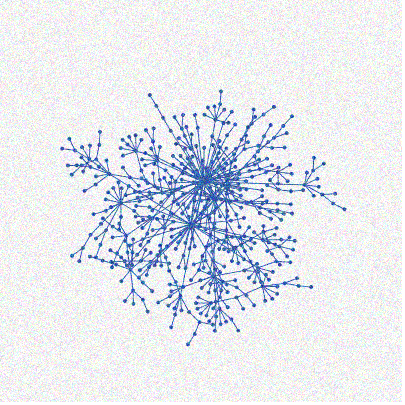}
  \caption{Scale-free network}\label{fig:sf-nw}
\end{subfigure}
\centering
\begin{subfigure}{.5\textwidth}
  \includegraphics[width=.9\linewidth]{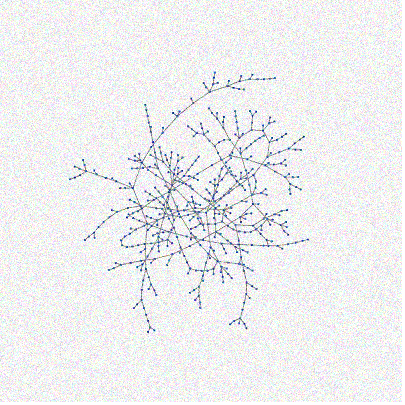}
  \caption{Random network}\label{fig:r-nw}
\end{subfigure}
\caption{Developed smart grid scenarios based on standard complex networks: Part(a) shows small-world constitutes of $500$ nodes, number of consumers and sources are selected randomly. Part(b) shows scale-free network with $500$nodes, a number of consumers and sources are selected randomly. Part(c) demonstrates random network consists of $500$nodes, a number of consumers and sources are selected randomly.}
\label{fig:crswsf}
\end{figure}


\begin{table}[H]
\begin{center}
\caption{Comparison on Random network}\label{tbl:ECRN}
\begin{tabular}{l | l}
\hline
\textbf{Technique} & \textbf{Time reduction (\%)} \\ \hline
RW vs Eigenvector & 55 \\ \hline
RW vs Betweenness & 35 \\ \hline
RW vs Closeness  & 54 \\ \hline
RW vs Degree & 56  \\ \hline
\end{tabular}
\end{center}
\end{table}

\begin{table}[H]
\begin{center}
\caption{Comparison on Scale-free network}\label{tbl:ECSF}
\begin{tabular}{l | l}
\hline
\textbf{Technique} & \textbf{Time reduction (\%)} \\ \hline
RW vs Eigenvector & 40.25 \\ \hline
RW vs Betweenness & 24 \\ \hline
RW vs Closeness  & 46.8 \\ \hline
RW vs Degree   & -82.5 \\ \hline
\end{tabular}
\end{center}
\end{table}

\begin{table}[H]
\begin{center}
\caption{Comparison on small-world network}\label{tbl:ECSW}
\begin{tabular}{l | l}
\hline
\textbf{Technique} & \textbf{Time reduction (\%)} \\ \hline
RW vs Eigenvector & 60 \\ \hline
RW vs Betweenness & 60 \\ \hline
RW vs Closeness  & 59.9 \\ \hline
RW vs Degree   & 59.8 \\ \hline
\end{tabular}
\end{center}
\end{table}

\begin{table}[H]
\caption{A comparative analysis of previous studies in the smart grid. We analyzed the previous study based on ABM, complex network, specification techniques ODD and DREAM. The comparative study confirms that there is no such specification study for ABM in the smart grid.}
\centering
\begin{tabular}{| l |p{2cm} |p{5.0cm} |p{1.0cm}|p{1.0cm} |p{1.0cm} |p{1.5cm}|}
\hline

\hline
Ref. & Author & Objective & ABM & CN & ODD & DREAM\\
\hline
\cite{ansari2016multi}	& Ansari &	MAS for reactive power management & \xmark & \xmark & \xmark & \xmark \\ \hline

\cite{wang2013adaptive}	& Zhang et al &	Adaptive strategy for energy trading & \cmark & \xmark & \xmark & \xmark \\\hline	

\cite{eriksson2015multiagent, ghorbani2016multiagent} &	Erikson et al. & Fault location and restoration in power system network	& \xmark	&\cmark	& \xmark	&\xmark \\\hline

\cite{samadi2016load} &	Samadi et al. &	Appliances scheduling in smart home & \xmark & \xmark & \xmark & \xmark \\\hline

\cite{tsai2017communication} &	Tsai et al.	& Communication among large-scale distributed consumers load	&\xmark & \xmark & \xmark & \xmark \\\hline

\cite{kremers2013multi} &	Kremers et al.	& Simple power system modeling with consumers and power generators	& \cmark	& \xmark & \xmark & \xmark \\\hline

\cite{li2015consensus}	& Li et al.	& Scheduling of flexible loads	& \cmark	& \xmark & \xmark & \xmark \\\hline

\cite{wang2016reinforcement} &	Wang et al. &	Modeling multiple microgrids	& \cmark	& \xmark & \xmark & \xmark \\\hline

\cite{kuznetsova2013reinforcement}	& Kuznetsova & Battery storage scheduling	& \cmark	& \xmark & \xmark & \xmark \\\hline

\cite{hinker2017novel} & Hinker	& A conceptual model for smart grid	& \xmark & \xmark & \cmark & \xmark \\\hline

\cite{wang2017synchronisation}	& Wang	& Frequency synchronization in power system & \xmark & \cmark & \xmark & \xmark \\\hline	

\cite{jia2012security} & Jia & Security analysis in power system & \xmark & \cmark & \xmark & \xmark \\\hline
\end{tabular}
\end{table}

\begin{table}[H]
\caption{Model specification follows ODD}
\centering
\begin{tabular}{| p{2.5cm} | p{3.0cm} | p{9.0cm} |}
\hline
Category &	Sub-category &	Our-model  \\\hline

\multirow{3}{*}{Overview} 
      & Purpose & Modeling smart grid using agent-based and complex network-based approach \\ 
      & Entities &  Producers, consumers, walkers\\  
      & Process & a hybrid centrality-based routing algorithm for end to end delivery from producers to consumers \\     
\hline
\multirow{3}{*}{Design concept} 
      & Basic principle &  a cognitive agent-based computing approach is better for modeling and simulation of the large scale power system. \\ 
      & Emergence & Computation time of end to end delivery from producers towards consumers \\  
      & Adaptation & Based on connected neighbors  \\   
      & Objective &  To measure how much time is taken while moving from one node to the other\\ 
      & Sensing & Check the state of the neighbor nodes \\  
      & Interaction & Local communication \\ 
      & Stochasticity & Random process \\ 
      & Obervation & Collect data about number of consumers, number of producers, number of nodes visited \\  
        
\hline
\multirow{3}{*}{Detail} 
      & Initialization & Complex network setup \\ 
      & Input data & External network setup files \\  
      & Sub-model &  Parameters: 
      \begin{itemize}
      \item number of nodes
      \item number of sources
      \item centrality-based routing
      \item average delivery rate calculation
\end{itemize}   \\     
\hline
\end{tabular}
\end{table}

\newpage
\section*{\refname}
\bibliography{sample}

\begin{thebibliography}{10}
\expandafter\ifx\csname url\endcsname\relax
  \def\url#1{\texttt{#1}}\fi
\expandafter\ifx\csname urlprefix\endcsname\relax\def\urlprefix{URL }\fi
\expandafter\ifx\csname href\endcsname\relax
  \def\href#1#2{#2} \def\path#1{#1}\fi

\bibitem{ellabban2016smart}
O.~Ellabban, H.~Abu-Rub, Smart grid customers' acceptance and engagement: An
  overview, Renewable and Sustainable Energy Reviews 65 (2016) 1285--1298.

\bibitem{epstein2008model}
J.~M. Epstein, Why model?, Journal of Artificial Societies and Social
  Simulation 11~(4) (2008) 12.

\bibitem{niazi2017towards}
M.~A. Niazi, Towards a novel unified framework for developing formal, network
  and validated agent-based simulation models of complex adaptive systems,
  Ph.D. thesis, University of Stirling (2011).

\bibitem{niazi2013complex}
M.~A. Niazi, Complex adaptive systems modeling: a multidisciplinary roadmap,
  Complex Adaptive Systems Modeling 1~(1) (2013) 1.

\bibitem{chassin2005evaluating}
D.~P. Chassin, C.~Posse, Evaluating north american electric grid reliability
  using the barab{\'a}si--albert network model, Physica A: Statistical
  Mechanics and its Applications 355~(2) (2005) 667--677.

\bibitem{batool2017modeling}
K.~Batool, M.~A. Niazi, Modeling the internet of things: a hybrid modeling
  approach using complex networks and agent-based models, Complex Adaptive
  Systems Modeling 5~(1) (2017) 4.

\bibitem{grimm2006standard}
V.~Grimm, U.~Berger, F.~Bastiansen, S.~Eliassen, V.~Ginot, J.~Giske,
  J.~Goss-Custard, T.~Grand, S.~K. Heinz, G.~Huse, et~al., A standard protocol
  for describing individual-based and agent-based models, Ecological modelling
  198~(1) (2006) 115--126.

\bibitem{niazi2012cognitive}
M.~A. Niazi, A.~Hussain, Cognitive agent-based computing-I: a unified framework
  for modeling complex adaptive systems using agent-based \& complex
  network-based methods, Springer Science \& Business Media, 2012.

\bibitem{grimm2010odd}
V.~Grimm, U.~Berger, D.~L. DeAngelis, J.~G. Polhill, J.~Giske, S.~F. Railsback,
  The odd protocol: a review and first update, Ecological modelling 221~(23)
  (2010) 2760--2768.

\bibitem{watts1998collective}
D.~J. Watts, S.~H. Strogatz, Collective dynamics of'small-world'networks,
  nature 393~(6684) (1998) 440.

\bibitem{barabasi1999mean}
A.-L. Barab{\'a}si, R.~Albert, H.~Jeong, Mean-field theory for scale-free
  random networks, Physica A: Statistical Mechanics and its Applications
  272~(1) (1999) 173--187.

\bibitem{barrenechea2004lattice}
G.~Barrenechea, B.~Beferull-Lozano, M.~Vetterli, Lattice sensor networks:
  Capacity limits, optimal routing and robustness to failures, in: Information
  Processing in Sensor Networks, 2004. IPSN 2004. Third International Symposium
  on, IEEE, 2004, pp. 186--195.

\bibitem{kleineberg2017collective}
K.-K. Kleineberg, D.~Helbing, Collective navigation of complex networks:
  Participatory greedy routing, Scientific Reports 7.

\bibitem{lin2016advanced}
B.~Lin, B.~Chen, Y.~Gao, K.~T. Chi, C.~Dong, L.~Miao, B.~Wang, Advanced
  algorithms for local routing strategy on complex networks, PloS one 11~(7)
  (2016) e0156756.

\bibitem{rekik2016geographic}
M.~Rekik, Z.~Chtourou, N.~Mitton, A.~Atieh, Geographic routing protocol for the
  deployment of virtual power plant within the smart grid, Sustainable Cities
  and Society 25 (2016) 39--48.

\bibitem{niazi2013modeling}
M.~A. Niazi, A.~Siddiqa, G.~Fortino, Modeling aids spread in social networks,
  in: German Conference on Multiagent System Technologies, Springer, 2013, pp.
  361--371.

\bibitem{niazi2014emergence}
M.~A. Niazi, Emergence of a snake-like structure in mobile distributed agents:
  an exploratory agent-based modeling approach, The Scientific World Journal
  2014.

\bibitem{hinker2017novel}
J.~Hinker, C.~Hemkendreis, E.~Drewing, S.~M{\"a}rz, D.~I.~H. Rodr{\'\i}guez,
  J.~M. Myrzik, A novel conceptual model facilitating the derivation of
  agent-based models for analyzing socio-technical optimality gaps in the
  energy domain, Energy.

\bibitem{jiang2017check}
Z.-Y. Jiang, J.-F. Ma, Y.-L. Shen, Check-in based routing strategy in
  scale-free networks, Physica A: Statistical Mechanics and its Applications
  468 (2017) 205--211.

\bibitem{niazi2009agent}
M.~Niazi, A.~Hussain, Agent-based tools for modeling and simulation of
  self-organization in peer-to-peer, ad hoc, and other complex networks, IEEE
  Communications Magazine 47~(3).

\bibitem{batool2014towards}
K.~Batool, M.~A. Niazi, S.~Sadik, A.~R.~R. Shakil, Towards modeling complex
  wireless sensor networks using agents and networks: a systematic approach,
  in: TENCON 2014-2014 IEEE Region 10 Conference, IEEE, 2014, pp. 1--6.

\bibitem{pradittasnee2017efficient}
L.~Pradittasnee, S.~Camtepe, Y.-C. Tian, Efficient route update and maintenance
  for reliable routing in large-scale sensor networks, IEEE Transactions on
  Industrial Informatics 13~(1) (2017) 144--156.

\bibitem{wang2017synchronisation}
C.~Wang, Synchronisation in complex networks with applications to power grids,
  Ph.D. thesis, University of Aberdeen (2017).

\bibitem{jia2012security}
Y.~Jia, Z.~Xu, S.-L. Ho, Z.~X. Feng, L.~L. Lai, Security analysis of smart
  grids-a complex network perspective.

\bibitem{guan2011routing}
Z.-H. Guan, L.~Chen, T.-H. Qian, Routing in scale-free networks based on
  expanding betweenness centrality, Physica A: Statistical mechanics and its
  applications 390~(6) (2011) 1131--1138.

\bibitem{yan2017efficient}
W.~yan Liu, X.~Li, J.~Li, B.~Liu, An efficient probability routing algorithm
  for scale-free networks, Chinese Journal of Physics 55~(3) (2017) 667--673.

\bibitem{oliveira2010centrality}
E.~M. Oliveira, H.~S. Ramos, A.~A. Loureiro, Centrality-based routing for
  wireless sensor networks, in: Wireless Days (WD), 2010 IFIP, IEEE, 2010, pp.
  1--5.

\bibitem{ansari2016multi}
J.~Ansari, A.~Gholami, A.~Kazemi, Multi-agent systems for reactive power
  control in smart grids, International Journal of Electrical Power \& Energy
  Systems 83 (2016) 411--425.

\bibitem{zhang2016agent}
X.~Zhang, A.~J. Flueck, C.~P. Nguyen, Agent-based distributed volt/var control
  with distributed power flow solver in smart grid, IEEE Transactions on Smart
  Grid 7~(2) (2016) 600--607.

\bibitem{eriksson2015multiagent}
M.~Eriksson, M.~Armendariz, O.~O. Vasilenko, A.~Saleem, L.~Nordstr{\"o}m,
  Multiagent-based distribution automation solution for self-healing grids,
  IEEE Transactions on Industrial Electronics 62~(4) (2015) 2620--2628.

\bibitem{ghorbani2016multiagent}
M.~J. Ghorbani, M.~A. Choudhry, A.~Feliachi, A multiagent design for power
  distribution systems automation, IEEE Transactions on Smart Grid 7~(1) (2016)
  329--339.

\bibitem{weng2017fault}
G.~Weng, F.~Huang, Y.~Tang, J.~Yan, Y.~Nan, H.~He, Fault-tolerant location of
  transient voltage disturbance source for dg integrated smart grid, Electric
  Power Systems Research 144 (2017) 13--22.

\bibitem{wang2013adaptive}
Z.~Wang, L.~Wang, Adaptive negotiation agent for facilitating bi-directional
  energy trading between smart building and utility grid, IEEE Transactions on
  Smart Grid 4~(2) (2013) 702--710.

\bibitem{tsai2017communication}
S.-C. Tsai, Y.-H. Tseng, T.-H. Chang, Communication-efficient distributed
  demand response: A randomized admm approach, IEEE Transactions on Smart Grid
  8~(3) (2017) 1085--1095.

\bibitem{kremers2013multi}
E.~Kremers, J.~G. de~Durana, O.~Barambones, Multi-agent modeling for the
  simulation of a simple smart microgrid, Energy Conversion and Management 75
  (2013) 643--650.

\bibitem{Chao2016}
H.-L. Chao, P.-A. Hsiung, A fair energy resource allocation strategy for micro
  grid, Microprocessors and Microsystems 42 (2016) 235--244.

\bibitem{li2015consensus}
Y.~Li, T.~Yong, J.~Cao, P.~Ju, J.~Yao, S.~Yang, A consensus control strategy
  for dynamic power system look-ahead scheduling, Neurocomputing 168 (2015)
  1085--1093.

\bibitem{wang2016reinforcement}
H.~Wang, T.~Huang, X.~Liao, H.~Abu-Rub, G.~Chen, Reinforcement learning in
  energy trading game among smart microgrids, IEEE Transactions on Industrial
  Electronics 63~(8) (2016) 5109--5119.

\bibitem{samadi2016load}
P.~Samadi, V.~W. Wong, R.~Schober, Load scheduling and power trading in systems
  with high penetration of renewable energy resources, IEEE Transactions on
  Smart Grid 7~(4) (2016) 1802--1812.

\bibitem{clausen2017agent}
A.~Clausen, A.~Umair, Y.~Demazeau, B.~N. J{\o}rgensen, Agent-based integration
  of complex and heterogeneous distributed energy resources in virtual power
  plants, in: International Conference on Practical Applications of Agents and
  Multi-Agent Systems, Springer, 2017, pp. 43--55.

\bibitem{shirzeh2015balancing}
H.~Shirzeh, F.~Naghdy, P.~Ciufo, M.~Ros, Balancing energy in the smart grid
  using distributed value function (dvf), IEEE Transactions on Smart Grid 6~(2)
  (2015) 808--818.

\bibitem{kuznetsova2013reinforcement}
E.~Kuznetsova, Y.-F. Li, C.~Ruiz, E.~Zio, G.~Ault, K.~Bell, Reinforcement
  learning for microgrid energy management, Energy 59 (2013) 133--146.

\end{thebibliography}
\bibliographystyle{elsarticle-num}

\end{document}